\documentclass[letterpaper,aps,prd,twocolumn,floats,preprintnumbers,footinbib,nosuperscriptaddress,showpacs]{revtex4}
\usepackage[pdftex]{hyperref}
\usepackage{amsmath,amssymb}
\usepackage{graphicx}
\usepackage{multirow} 
\usepackage{bigstrut} 
\usepackage{xcolor}

\begin{document}

\preprint{OUTP-0922P}
\title{On cosmic ray acceleration in supernova remnants and the
  FERMI/PAMELA data}
\author{Markus Ahlers}
\affiliation{Rudolf Peierls Centre for Theoretical Physics, University
  of Oxford, Oxford OX1 3NP, UK}
\author{Philipp Mertsch}
\affiliation{Rudolf Peierls Centre for Theoretical Physics, University
  of Oxford, Oxford OX1 3NP, UK}
\author{Subir Sarkar}
\affiliation{Rudolf Peierls Centre for Theoretical Physics, University
  of Oxford, Oxford OX1 3NP, UK}
  

\begin{abstract}
  We discuss recent observations of high energy cosmic ray positrons
  and electrons in the context of hadronic interactions in supernova
  remnants, the suspected accelerators of galactic cosmic
  rays. Diffusive shock acceleration can harden the energy spectrum of
  secondary positrons relative to that of the primary protons and
  electrons and thus explain the rise in the positron fraction
  observed by PAMELA above 10~GeV. We normalize the hadronic
  interaction rate by holding pion decay to be responsible for the
  gamma-rays detected by HESS from some SNRs. By simulating the
  spatial and temporal distribution of SNRs in the Galaxy according to
  their known statistics, we are able to then fit the electron (plus
  positron) energy spectrum measured by Fermi. It appears that IceCube
  has good prospects for detecting the hadronic neutrino fluxes
  expected from nearby SNRs.
 
\end{abstract}

\pacs{98.70.Sa, 96.50.sb, 98.58.Mj}

\maketitle

\section{Introduction}

Recently, the PAMELA collaboration has published data on the positron
fraction, $J_{e^+} / (J_{e^-} + J_{e^+})$, in galactic cosmic rays
(GCR) which is seen to \emph{increase} between $\sim 10$ and 100~GeV
\cite{Adriani:2008zr}, in contrast to the prediction of the standard
GCR propagation model \cite{Moskalenko:1997gh}. The model assumes that
the positrons are secondaries created via interactions of GCR protons
and nuclei with interstellar matter, hence their spectrum should be
softer than that of primary electrons and the positron fraction should
thus decrease with energy \cite{Serpico:2008te}. The combined
differential flux of GCR electrons and positrons, $(J_{e^-} +
J_{e^+})$, has also been measured with the Fermi Large Area Telescope
(LAT) and is approximately fitted by a $E^{-3}$ power-law in energy up
to $\sim 1$~TeV \cite{Abdo:2009zk}. Measurements by the HESS
collaboration ~\cite{Collaboration:2008aaa,Aharonian:2009ah} show
significant steepening of the spectrum beyond $\sim 1$~TeV, while
agreeing well with the Fermi data at lower energies. Although the
Fermi LAT data does not confirm the sharp feature claimed earlier by
the ATIC collaboration \cite{Chang:2008zzr}, there does appear to be a
small excess flux above $\sim 100$~GeV in comparison with standard GCR
propagation models~\cite{Moskalenko:1997gh,Grasso:2009ma}.

Both these experimental findings have generated a lot of interest
because they may be an indirect signature of dark matter
particles. Annihilation or decay of galactic dark matter can produce
electrons and positrons with a spectrum considerably harder than that
of primary electrons. Besides the fine-tuning challenges such models
face \cite{Grasso:2009ma,Bergstrom:2009ib}, other cosmic ray data
provide important constraints. The antiproton-to-proton ratio observed
by PAMELA~\cite{Adriani:2008zq} is in good agreement with the
prediction of secondary production by GCRs and thus rules out most
dark matter annihilation/decay models which have hadronic final
states. Even purely leptonic annihilation channels are strongly
constrained by the Galactic synchrotron radio
background~\cite{Bertone:2008xr} and by the Galactic gamma-ray
background~\cite{Cirelli:2009vg}. In fact pulsars may produce a hard
spectrum of electron-positron pairs in the right energy range to
explain both the positron flux anomaly and the observed electron flux
\cite{AharonianPULSAR,Hooper:2008kg}.

It has long been believed \cite{Ginzburg:1990sk} that GCRs are
generated by diffusive shock acceleration (DSA)
\cite{Blandford:1987pw,Malkov:2001} in supernova remnants (SNRs)
\cite{Reynolds:2008}. Hadronic interactions of the accelerated protons
will create $\pi^\pm$ (and $\pi^0$) which then decay to yield
secondary $e^{\pm}$ and neutrinos (and $\gamma$-rays). It has been
suggested that the {\it acceleration} of the secondary positrons in a
nearby SNR shock wave may be responsible for the PAMELA anomaly
\cite{Blasi:2009hv}. The fraction of secondary $e^+$ which are
accelerated increases with energy, so their final spectrum is {\em
  harder} than the injected spectrum. This effect had been noted
earlier as a general expectation for the secondary-to-primary ratio in
the presence of stochastic Fermi acceleration
\cite{Eichler:1980hw,Cowsik:1980ApJ}. A similar effect is then
predicted at higher energies for both antiprotons \cite{Fujita:2009wk,Blasi:2009bd}
and for secondary nuclei such as boron \cite{Mertsch:2009ph}. These
predictions will be tested soon with data from PAMELA and the
forthcoming AMS-02 mission \cite{AMS02}.

It is interesting to ask whether this model can account also for the
absolute fluxes of $e^-$ (and $e^+$) in GCR at energies $\gtrsim
100$~GeV \cite{Blasi:2009hv}. This is rather sensitive to the assumed
spatial distribution of the sources so in this paper we consider a
realistic distribution of SNRs based on astronomical data
(Sec.~\ref{sec:SourceDistribution}). Previously the flux of secondary
$e^-$ and $e^+$ in the sources has been normalized with respect to the
primary electrons in an {\it ad hoc} fashion
\cite{Blasi:2009hv}. Instead, we exploit the hadronic origin of these
secondaries and normalize using the $\gamma$-ray fluxes (assumed to be
from $\pi^0$ decay) detected from known SNRs by HESS. We can thus fix
the only free model parameter by fitting the total $e^- + e^+$ flux to
Fermi LAT and HESS data (Sec.~\ref{sec:Spectra}). The $e^+$ fraction
is then \emph{predicted} up to TeV energies and provides a good match
to PAMELA data (Sec.~\ref{sec:Results}). Having constrained the
distribution of the closest SNRs via the measured $e^-$ and $e^+$
spectra, we present an example of a likely source distribution in
order to illustrate that there are good prospects for IceCube to
detect neutrinos from nearby SNR. A consistent picture thus emerges
for all presently available data in the framework of the standard
DSA/SNR origin model of GCR. However there remain some open issues and
grounds for concern which we discuss at the end
(Sec.~\ref{sec:DiscussionSummary}).

\begin{figure*}[t]\centering
\begin{minipage}[t]{\columnwidth}\centering
\includegraphics[width=0.85\columnwidth]{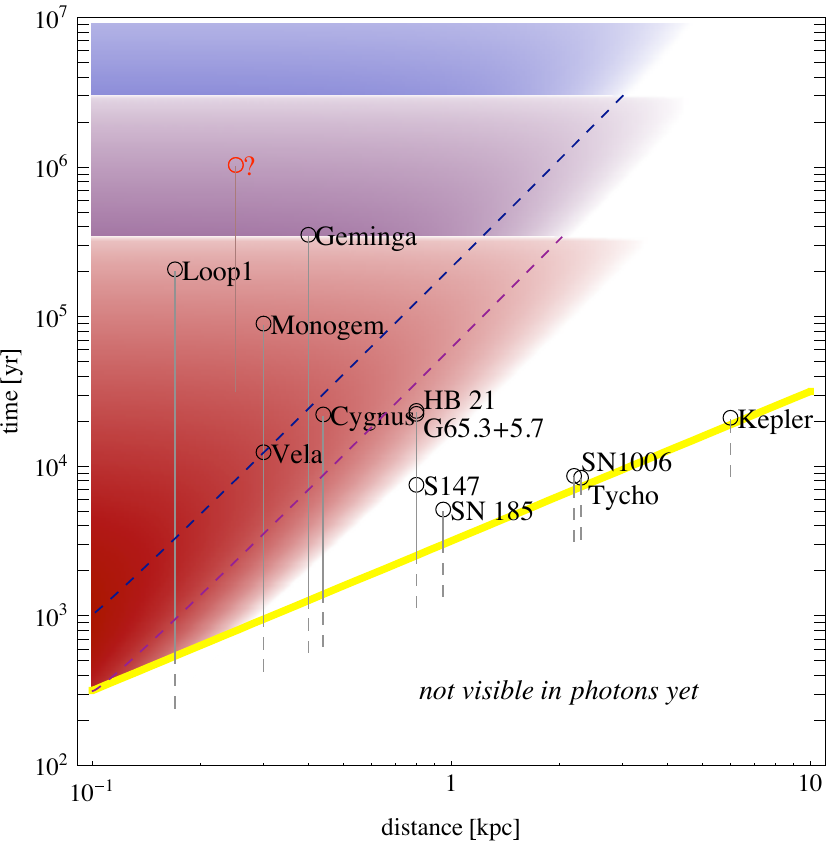}
\end{minipage}\hfill
\begin{minipage}[t]{\columnwidth}\centering
\includegraphics[width=0.85\columnwidth]{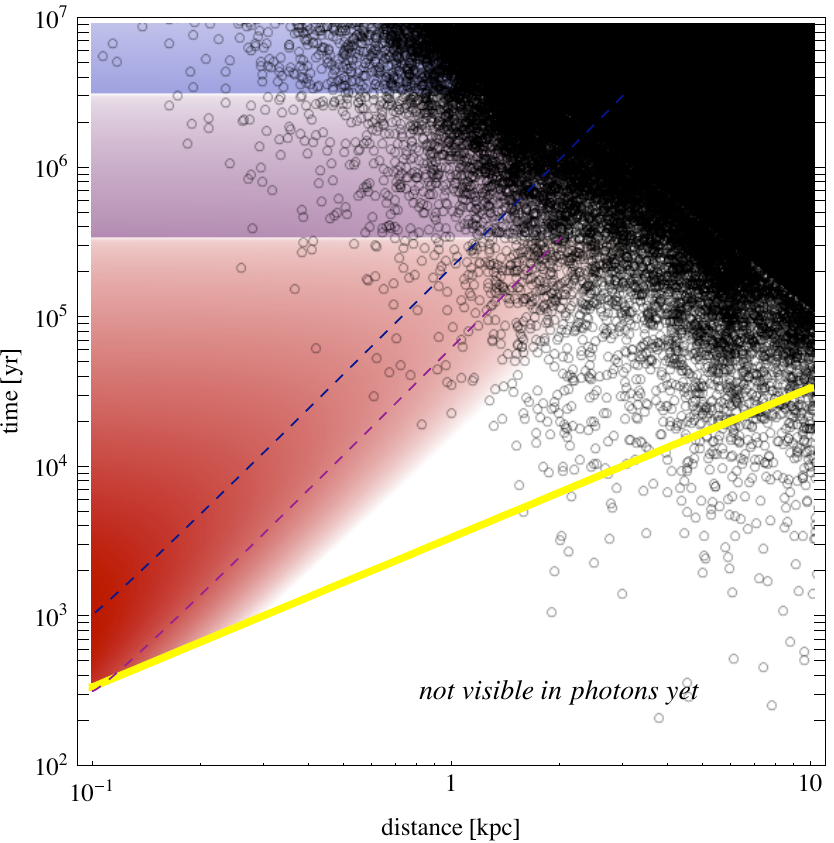}
\end{minipage}
\caption[]{Distance-time diagram for nearby SNRs (after
  Ref.~\cite{Swordy:2003ds}). {\bf Left:} The open circles mark
  supernova events and the world-lines of the discovered remnants are
  indicated. The thick yellow line is our past light-cone; all events
  lying on it, {\it e.g.}, the SNR world-lines touching it, can be
  observed presently. The blue, purple and red shadings (top to
  bottom) show the relative contribution of sources to the diffuse
  $e^-$ and $e^+$ flux observed at Earth at 10, 100 and 1000 GeV,
  respectively. The open red circle is an example of a hypothetical
  supernova whose remnant is too old to be visible any longer but
  which might still be contributing to the diffuse $e^-$ and $e^+$
  flux. {\bf Right:} A distance-time diagram for hypothetical nearby
  SNRs. The open black circles are an example of a possible
  ``history'' of supernovae (the wordlines have been suppressed) as
  simulated by our Monte Carlo calculation (see
  Section~\ref{sec:SourceDistribution}).}
\label{fig:distance-time-diagram}
\end{figure*}

\section{Diffusion model and source distribution}
\label{sec:SourceDistribution}

The diffusive transport of high energy $e^-$ and $e^+$ in the Galaxy
is governed by the equation \cite{Ginzburg:1990sk},
\begin{equation}
\label{eqn:diffusion}
\frac{\partial n_\pm}{\partial t} = \nabla\left(D_\text{GCR}\nabla n_\pm\right)
+ \frac{\partial}{\partial E} \left(b n_\pm\right) + Q_\pm \,,
\end{equation}
where $n_\pm \text{d}E \equiv n_\pm (r, t, E) \text{d}E$ denotes the
particle density of $e^+$ and $e^-$ with energy in $[E, E +
\text{d}E]$. The spatial diffusion coefficient is assumed not to
depend on the position in the Galaxy but only on energy: $D_\text{GCR}
(E) = D_0 E^\delta$. The energy loss rate of GCR $e^-$ and $e^+$
through synchrotron radiation in Galactic magnetic fields and inverse
Compton scattering on the CMB and interstellar radiation backgrounds
is parametrized as $b (E) = b_0 E^2$. Finally, $Q_\pm$ denotes the
injection of electrons and positrons from both (possible) primary and
secondary sources. In the majority of previous calculations, in
particular the GALPROP code \cite{Moskalenko:1997gh} in its
conventional setup, the distribution of sources is assumed to be
continuous. However at energies when the diffusion length $\ell$
becomes smaller than the distance to the closest source (similar to
the average distance between sources for a homogeneous distribution)
the fact that the sources are discrete should become important. This
effect on GCR electrons was first pointed out in
Ref.~\cite{Cowsik:1979} and has later been considered in more detail
\cite{Pohl:1998ug,Kobayashi:2003kp,Cowsik:2009ec}. It has also been
studied using an extended version of GALPROP \cite{Swordy:2003ds}.

Assuming that all the electrons and positrons are released instantly
at the end of the Sedov-Taylor phase of expansion when the SNR becomes
radiative, the flux of $N$ sources at distances $r_i$ and times $t_i$
is given by the sum over the corresponding Green's functions (see
Appendix~\ref{Green}) for the particle density (times the usual ``flux
factor'', $c/4\pi$, for an isotropic population of relativistic
particles):
\begin{align}
J_N (E) &= \frac{c}{4\pi} \sum_{i=1}^N G_\text{disk} (E , r_i , t_i).
\label{eqn:flux}
\end{align}
The spatial distribution and temporal history of GCR sources, $\{r_i,
t_i\}$, is not known {\it a priori} but can be modelled for an assumed
source class {\it e.g.}, SNRs. At low energies, the GCR diffusion
length is long enough such that the approximation of a continuous
source density is acceptable. For energies above some hundreds of GeV,
however, the fluctuations introduced by the discreteness of the
sources cannot be neglected any longer.

\begin{figure*}[t]
\begin{minipage}[t]{\columnwidth}\centering
\includegraphics[width=0.85\columnwidth]{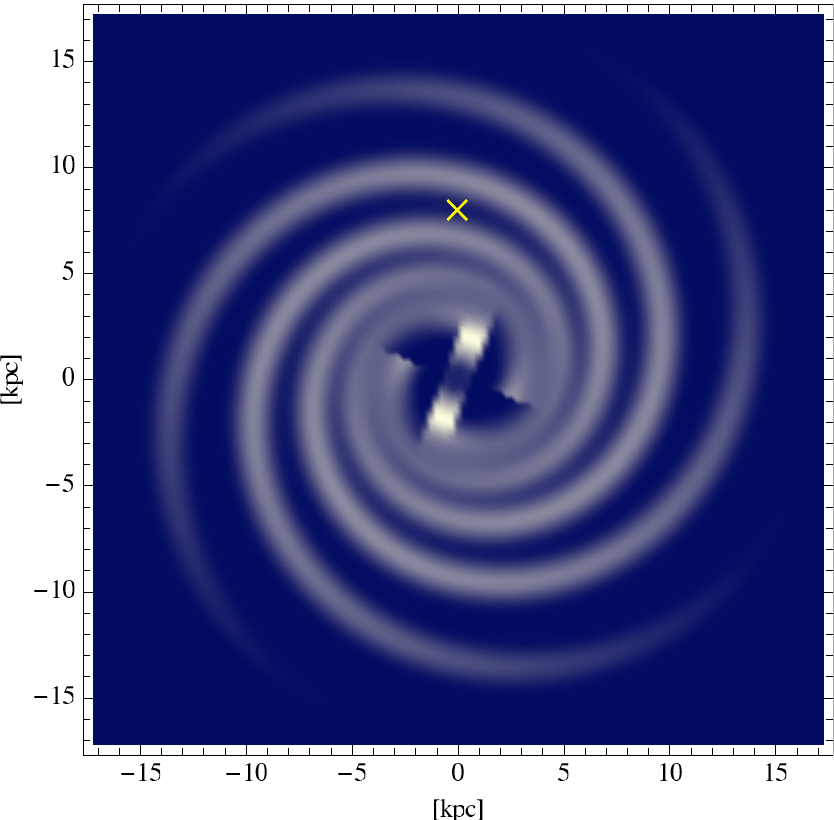}
\end{minipage}\hfill
\begin{minipage}[t]{\columnwidth}\centering
\includegraphics[width=0.85\columnwidth]{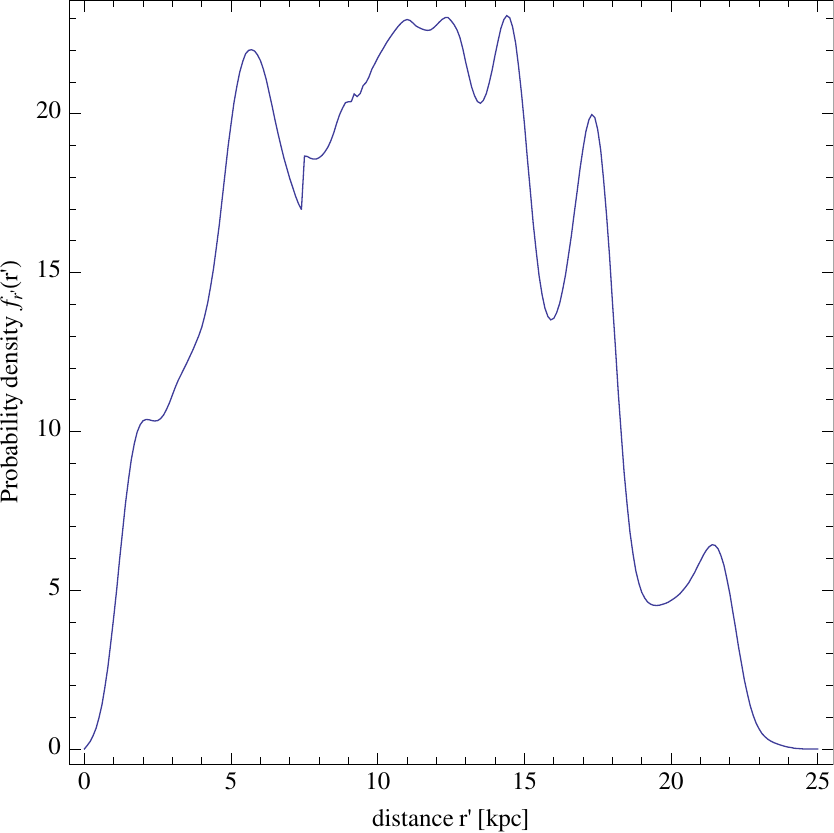}
\end{minipage}
\caption{{\bf Left:} The assumed distribution of SNRs in the Galaxy;
  the cross denotes the position of the Sun in between two spiral
  arms. {\bf Right:} The probability density for the distance of a SNR
  from the Sun.}
\label{fig:SNR_density&spiral_structure}
\end{figure*}

Some authors \cite{Kobayashi:2003kp,Shaviv:2009bu} have assumed a
continuous distribution of sources for distances beyond a few hundred
parsecs, supplemented by a set of known SNRs for smaller
distances. This approach is however biased by the choice of young,
nearby sources which have been detected in radio and/or X-rays. Older
sources may not be visible in photons any longer but still be
contributing to the GCR electron flux, see
Fig.~\ref{fig:distance-time-diagram}. We note that the effect of this
incomplete assumed source distribution is a dip in the electron flux
seen in both analyses \cite{Kobayashi:2003kp,Shaviv:2009bu}, although
at different energies because of the different diffusion model
parameters chosen.

Determining the complete distribution of sources in our vicinity ({\it
  i.e.}~up to a few kpc) from observations seems challenging. However
it turns out that we do not need to know the exact distribution in
order to calculate the $e^+$ flux but require only a limited amount of
information, most of which is encoded already in the dominant $e^-$
flux. By including the recent measurements by Fermi LAT
\cite{Abdo:2009zk} and HESS
\cite{Collaboration:2008aaa,Aharonian:2009ah} of the total $e^- + e^+$
flux in the energy region of interest, we have sufficient information
at hand to make a prediction for the positron fraction under the
assumption that the additional positrons originate in the same
sources.

We perform a Monte Carlo calculation by considering a large number of
random distributions of sources drawn from a probability density
function that reflects our astronomical knowledge about the
distribution of SNRs in the Galaxy. The better the flux of $e^-$ and
$e^+$ from such a ``history'' of sources reproduces the measured
fluxes, the closer is the underlying distribution of sources likely to
be to the actual one. Of course all SNRs are not the same, however
variations of the source parameters would only introduce additional
fluctuations into the fluxes without altering their average. We can
choose the best ``fit'' to the data and thus determine the $e^+$ flux.

\begin{table*}[!th]
\begin{minipage}[t]{\textwidth}\setlength{\tabcolsep}{4pt}
  \begin{center}
    \begin{tabular}{l | l | | r@{$\:\pm\:$} l  r@{$\:\pm\:$} l c | c | c | r }
      \hline\hline
         Source
      & Other name(s)
      & \multicolumn{2}{c}{$\Gamma$} 
      & \multicolumn{2}{c}{$J_\gamma^0 \div 10^{-12}$}
      & $E_\text{max}$ 
      & $d$
      & $Q_{\gamma}^0 \div 10^{33}$
      & Ref.
      \\
      & 
      &\multicolumn{2}{c}{ } 
      &\multicolumn{2}{c}{\hspace{-0.2cm} 
        \footnotesize [$({\rm cm}^{2}\,{\rm s}\,{\rm TeV})^{-1}$]}  
      & {\footnotesize [TeV]}
      & {\footnotesize [kpc]}
      & {\footnotesize [$({\rm s}\,{\rm TeV})^{-1}$]}
      & 
      \\
      \hline
      HESS~J0852$-$463 & RX J0852.0-4622 (Vela Junior) 
      & 2.1 & 0.1 & 21 & 2 & $> 10$ & 0.2 & 0.10 & \cite{Aharonian:2005} \\
      HESS~J1442$-$624 &RCW 86, SN 185 (?) 
      & 2.54 & 0.12 & 3.72 & 0.50 & $\gtrsim$ 20 & 1 & 0.46 
      & \cite{Aharonian:2008nw}\\
      HESS~J1713$-$381 & CTB 37B, G348.7+0.3 
      & 2.65  & 0.19  & 0.65 & 0.11 & $\gtrsim$ 15 & 7 & 3.812 
      & \cite{Aharonian:2008ka}\\
      HESS~J1713$-$397 & RX J1713.7-3946, G347.3-0.5 
      & 2.04 & 0.04 & 21.3 & 0.5 & 17.9 $\pm$ 3.3 & 1 & 2.55 
      & \cite{Aharonian:2006, Aharonian:2006ws} \\
      HESS~J1714$-$385 & CTB 37A 
      &  2.30 & 0.13 & 0.87 & 0.1 & $\gtrsim$ 12 & 11.3 & 13.3 
      & \cite{Aharonian:2008km} \\
      HESS~J1731$-$347 & G 353.6-07 
      & 2.26 & 0.10 & 6.1  & 0.8 & $\gtrsim$ 80 & 3.2 & 7.48 
      & \cite{Aharonian:2008, Tian:2008tr}  \\
      HESS~J1801$-$233\footnote{We assume that W 28 powers only the emission from J1801$-$233 (and not the nearby J1800$-$240 A, B and C).} 
      & W 28, GRO J1801-2320 
      & 2.66  & 0.27  & 0.75 & 0.11 & $\gtrsim$ 4 & 2 & 0.359 
      & \cite{Aharonian:2008fp} \\
      HESS~J1804$-$216\footnote{W30 is taken to be the origin of the VHE emission \cite{Fatuzzo:2006ua}.}  & W 30, G8.7-0.1 
      & 2.72 & 0.06  & \multicolumn{2}{c}{5.74} & $\gtrsim$ 10 & 6 & 24.73 
      & \cite{Aharonian:2005kn} \\
      HESS~J1834$-$087 &W 41, G23.3-0.3 
      & 2.45 & 0.16 & \multicolumn{2}{c}{2.63} & $\gtrsim$ 3 & 5 & 7.87 
      & \cite{Aharonian:2005kn} \\
      MAGIC J0616+225 & IC 443 
      & 3.1 & 0.3 & \multicolumn{2}{c}{0.58} & $\gtrsim 1 $ & 1.5 & 0.156 
      & \cite{Albert:2007tr} \\
      Cassiopeia A 
      & & 2.4 & 0.2 & 1.0 & 0.1 & $\gtrsim 40$ &  3.4 & 1.38 
      & \cite{Albert:2007wz}\footnote{Cas A was first detected by HEGRA \cite{Aharonian:2001}.} \\
      J0632$+$057 & Monoceros 
      & 2.53 & 0.26 & 0.91 & 0.17 & N/A & 1.6 & 0.279 
      & \cite{Fiasson:2007bk} \\
      \hline
      \multicolumn{2}{l ||}{Mean}& \multicolumn{2}{c}{$\sim 2.5$} & \multicolumn{2}{c}{}  
      & $\gtrsim 20$ & & $\sim 5.2$ & \\  
      \hline
      \multicolumn{2}{l ||}{Mean, excluding sources with $\Gamma > 2.8$}& \multicolumn{2}{c}{$\sim 2.4$} & \multicolumn{2}{c}{}  
      & $\gtrsim 20$ & & $\sim 5.7$ & \\  
      \hline
      \multicolumn{2}{l ||}{Mean, excluding sources with $\Gamma > 2.6$}& \multicolumn{2}{c}{$\sim 2.3$} & \multicolumn{2}{c}{}  
      & $\gtrsim 20$ & & $\sim 4.2$ & \\  
      \hline\hline      
    \end{tabular}
    \caption{Summary of spectral parameters for SNRs detected in
      $\gamma$-rays from a power-law fit to the spectrum, $J_\gamma =
      J_\gamma^0 (E/\text{TeV})^{-\Gamma}$, with an exponential
      cut-off at $E_\text{max}$ in the case of HESS~J1713.7-397. The
      errors shown are statistical only --- the systematic error is
      conservatively estimated to be 20\% on the flux $J_\gamma$ and
      $\pm 0.2$ on the spectral index $\Gamma$. Also shown is the
      estimated distance $d$ and the injection rate $Q_\gamma^0$
      derived from Eq.~(\ref{eqn:GammaLumi}).}
    \label{tbl:GammaRays}
  \end{center}
  \end{minipage}
\end{table*}

The smoothed radial distribution of SNRs in the Galaxy is well
modelled by \cite{Case:1998qg}:
\begin{equation}
f (r) = A \sin{\left(\frac{\pi r}{r_0} + \theta_0\right)} \text{e}^{-\beta r},
\label{eqn:SNR_density}
\end{equation}
where $A = 1.96\,\text{kpc}^{-2}$, $r_0 = 17.2\,\text{kpc}$, $\theta_0
= 0.08$ and $\beta = 0.13$. To obtain a realistic probability density
for the distance between the Earth and a SNR we have to also take into
account the spiral structure of the Galaxy. We adopt a logarithmic
spiral with four arms of pitch angle $12.6^\circ$ and a central bar of
6 kpc length inclined by $30^\circ$ with respect to the direction Sun
- Galactic centre \cite{Vallee:2005}. The density of SNRs is modelled
by a Gaussian with 500 pc dispersion for each arm
\cite{Pohl:1998ug}. The resulting distribution $g (r, \phi)$ (see left
panel of Fig.~\ref{fig:SNR_density&spiral_structure}) has been
normalized with respect to azimuth in such a way that the above radial
distribution (\ref{eqn:SNR_density}) is recovered. To obtain the
probability density for the distances we transform to the coordinates
$(r', \phi')$ centered on the Sun. As the $e^-$ and $e^+$ fluxes are
assumed to be isotropic, we can average over the polar angle $\phi'$,
such that the probability density $f_{r'}$ depends only on the
distance $r'$ to the source,
\begin{equation}
 f_{r'} (r') = \frac{1}{2 \pi} \int_0^{2\pi} \text{d}\phi' r' g (r (r', \phi'), 
 \phi (r', \phi')).  
\label{eqn:dF}
\end{equation}
This function is shown in the right panel of
Fig.~\ref{fig:SNR_density&spiral_structure}.

We assume that the sources are uniformly distributed in time, {\it
  i.e.} their probability density $f_t (t)$ is
\begin{equation}
 f_t (t) = \left\{\begin{array}{ll} 1/t_\text{max} & \text{for } 
 0 \leq t \leq t_\text{max}, \\ 0 & \text{otherwise}, \end{array}\right.
\end{equation}
with $t_\text{max}$ standing for the earliest time considered, which
is related to the minimum energy for which our calculation is valid
through:
\begin{equation}
t_\text{max} = \left(b E_{\text{min}}\right)^{-1}.
\end{equation}
The total number $N$ of sources that are needed in the Monte Carlo to
reproduce the (observed) number $\mathcal{N}~\simeq~300$ of SNRs
active in the galaxy at any given time depends on the average lifetime
of a SNR, $\tau_\text{SNR}$, which is suggested to be $\sim 10{^4}$~yr
\cite{Reynolds:2008}, hence
\begin{align}
  N = 3 \times 10^6 \left(\frac{\mathcal{N}}{300}\right) 
   \left(\frac{t_\text{max}}{10^8\,\text{yr}}\right) 
   \left(\frac{\tau_\text{SNR}}{10^4\,\text{yr}}\right)^{-1}.
\end{align}
%

\section{Fitting the $e^+ + e^-$ spectra}
\label{sec:Spectra}

\begin{figure}[!b]
\centering
\includegraphics[width=1.0 \columnwidth]{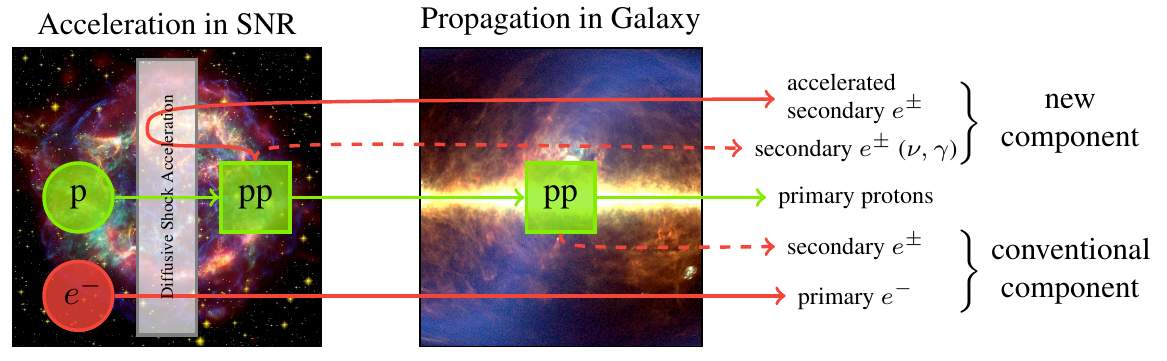}
\caption[]{Schematic description of contributions to the galactic
  cosmic rays observed at Earth in the present framework.}
\label{fig:SNR_mechanism}
\end{figure}

A schematic description of the present framework is shown in
Fig.~\ref{fig:SNR_mechanism}. Cosmic rays are shock accelerated in
SNRs and then diffuse through the Galaxy to Earth undergoing
collisions with interstellar matter {\it en route} and creating
secondary $e^+$. As discussed, the ratio of the secondary $e^+$ to the
primary $e^-$ from the sources should {\it decrease} with energy, in
contrast to the behaviour seen by PAMELA. We follow
Ref.~\cite{Blasi:2009hv} in explaining this by invoking a new
component of $e^+$ which is produced through cosmic ray interactions
in the SNRs, and then shock {\it accelerated}, thus yielding a harder
spectrum than that of their primaries. We discuss these components in
turn below and calculate their relative contributions by normalising
to the $\gamma$-ray flux from the SNRs, which provides an independent
measure of the hadronic interactions therein.

\subsection{Primary electrons}
\label{sec:PrimaryElectrons}

The radio and X-ray emission observed from SNRs is interpreted as
synchrotron radiation of electrons accelerated up to energies of
${\cal O}(100)$~TeV \cite{Reynolds:2008}. The spectrum of this
radiation then determines the spectrum of the underlying relativistic
electrons. Moreover the theory of diffusive shock acceleration
\cite{Blandford:1987pw,Malkov:2001} predicts similar spectra for the
accelerated protons and nuclei as for the electrons. If the
$\gamma$-ray emission observed by HESS from a number of identified
SNRs is assumed to be of hadronic origin, we can use the
measured spectra to constrain both the relativistic proton and
electron population.

Table~\ref{tbl:GammaRays} shows a compilation of $\gamma$-ray sources
observed by HESS that have been identified as SNRs. We have included
all identified shell-type SNRs and strong SNR candidates in the HESS
source catalogue \cite{HESS:SourceCatalogue}, and also added the SNRs
IC 443, Cassiopeia A and Monoceros. Actually it is not clear that the
acceleration of secondaries does occur in all the SNRs considered,
especially when the $\gamma$-ray emission is associated with a
neighbouring molecular cloud rather than coming from the vicinity of
the shock wave. In fact the $\gamma$-rays could equally well be due to
inverse-Compton scattering by the relativistic electrons responsible
for the observed synchroton radio and X-ray emission. Therefore, we
have considered three possibilities --- including all sources implies
a mean power-law spectral index for the protons of $\langle \Gamma
\rangle = 2.5$, while excluding steep spectrum sources with $\Gamma >
2.8$ gives $\langle \Gamma \rangle = 2.3$ and excluding sources with
$\Gamma > 2.6$ yields $\langle \Gamma \rangle = 2.4$. In the following
we adopt the central value, $\Gamma =2.4$, for the electron population
too, unless stated otherwise. This requires a compression factor of $r
\leq 3.3$ in contrast to the value of $r = 4$ expected for a strong
shock, so there is clearly some tension between the DSA theory and
observations. This can possibly be resolved if we consider only a
subset of the SNRs in Table 1 to be hadronic accelerators, or if the
$\gamma$-ray spectrum is steepened e.g. by the onset of an exponential
cutoff in the electron spectrum. Our model assumptions are intimately
connected to the production of neutrinos, the detection of which will
therefore provide an independent test as we discuss later. In this
work we adopt a cut-off of $E_\text{cut} \simeq 20$~TeV which is
consistent with DSA theory \cite{Reynolds:2008}. The source spectrum
of primary electrons is then:
\begin{equation}
  R_{e^-} = R_{e^-}^0 \left(\frac{E}{\text{GeV}}\right)^{-\Gamma} 
   \text{e}^{-E/E_\text{cut}}.
\end{equation}
The normalisation $R_{e^-}^0$ is determined by fitting the electron
flux at Earth resulting from our Monte Carlo computation to the
preliminary measurement by PAMELA at 10~GeV~\cite{Mocchiutti:2009}; the
secondary fluxes can be neglected for this normalisation. We find
$R_{e^-}^0 = 1.8 \times 10^{50} \,\text{GeV}^{-1}$ for $\Gamma = 2.4$
which corresponds to a total injection energy of
\begin{align}
\int_{1 \,\text{GeV}}^{20 \,\text{TeV}} \text{d}E \,E \,R_{e^-}(E) 
\simeq 7 \times 10^{47} \,\text{erg}.
\end{align}
This compares well to the value of $9.2 \times 10^{47}$ erg
said to be required to power the GCR electrons \cite{Reynolds:2008}.  

Solar modulation which is important below $\sim 10$~GeV, has been
accounted for using the force field approach \cite{Gleeson:1968}, with
a charge-independent potential of $\phi = 600 \, \text{MV}$.  However,
our simple model ignores convection and (re)acceleration in the
interstellar medium which become important below $\sim 5
\,\text{GeV}$, hence the electron flux cannot be predicted at lower
energies. The primary $e^-$ fluxes as measured on Earth for 30
different source configurations are shown in the top panel of
Fig.~\ref{fig:E3J}. With an injection power-law index $\Gamma \simeq
2.4\pm0.1$ as required for consistency with the $\gamma$-ray data,
there clearly is a deficit at high energies compared to the $e^+ +
e^-$ flux measured by Fermi LAT and HESS.

\subsection{Secondary electrons and positrons}

Positrons in GCR are generally assumed to be of purely secondary
origin, arising through the decay of pions and kaons produced in the
interactions of GCR protons (and nuclei) with the interstellar medium
(ISM) \cite{Moskalenko:1997gh}. The neutral pions decay into
$\gamma$-rays which then contribute to, if not dominantly constitute,
the Galactic $\gamma$-ray background. The charged pions on the other
hand decay into neutrinos and muons, the latter subsequently decaying
into electrons and positrons. Assuming that spatial and temporal
variations in the GCR proton flux $J_\text{p}$ and the ISM gas density
$n_\text{ISM}$ are small, the source density of these secondary
background $e^-$ and $e^+$ is also homogeneous, both in space and in
time:
\begin{equation}
 q^\text{ISM}_\pm = n_\text{ISM}\, c \int_{E_\text{thr}}^\infty 
 \text{d}E' \frac{4\pi}{\beta c} J_\text{p} (E')  
 \frac{\text{d}\sigma_{\text{pp} \rightarrow \text{e}^\pm + X}}{\text{d}E},
\end{equation}
where $\text{d}\sigma_{\text{pp} \rightarrow \text{e}^{\pm} +
  X}/\text{d} E$ is the partial differential cross-section for $e^\pm$
production and $\beta \simeq 1$ is the velocity of the GCR. We can
then integrate the Green's function for a single source over space and
time to calculate
\begin{equation}
 J_{\pm}(E) \simeq \frac{c}{4\pi} \frac{1}{|b(E)|} \int_E^{\infty} \text{d}E' 
 q^\text{ISM}_\pm (E') \frac{2h}{\ensuremath{\ell_\text{cr}}} \chi 
 \left(0, \frac{\ell}{\ensuremath{\ell_\text{cr}}}\right),
\end{equation}
where $\chi$ and $\ell_{\text{cr}}$ are defined by Eq.~(\ref{functionchi}) of
Appendix~\ref{Green}, $\ell$ is the diffusion length defined by
Eq.~(\ref{difflength}), and $h \sim 0.1$~kpc is the height of the
Galactic disk.

We calculate the flux of secondary background $e^-$ and $e^+$ from the
Solar-demodulated flux of GCR protons as derived from the BESS data
\cite{Shikaze:2006je} and model the cross-sections according to
Ref.~\cite{Kamae:2006bf}. The contribution from kaon decay is
subdominant and is therefore neglected. The presence of He both in
GCRs and in the ISM is taken into account by multiplying the proton
contribution by a factor of 1.2. Our results are in good agreement
with Ref.~\cite{Delahaye:2008ua}, taking into account the different
diffusion model parameters and keeping in mind that convection and
reacceleration have been neglected here. These fluxes are shown
(dashed line) in the middle panel of Fig.~\ref{fig:E3J} and are
clearly a subdominant component which cannot acount for the deficit at
high energies.

Moreover, the positron flux is {\it falling} at all energies whereas
the PAMELA data \cite{Adriani:2008zr} clearly show a {\it rise} above
a few GeV. One way this can be resolved is if there is a dip in the
electron spectrum between $\sim 10$ and 100 GeV. It has been suggested
that Klein-Nishina corrections to the Thomson cross section for
inverse Compton scattering \cite{Stawarz:2009ig} or inhomogeneities in
the distribution of sources \cite{Shaviv:2009bu} can produce such a
dip. However the former would require a rather enhanced interstellar
background light (IBL) field \cite{Stawarz:2009ig}, while the latter
calculation \cite{Shaviv:2009bu} assumes an incomplete source
distribution (see Sec.~\ref{sec:SourceDistribution}) and moreover
adopts diffusion model parameters quite different from those derived
from the measured nuclear secondary-to-primary ratios
\cite{Strong:2007nh} and the measured Galactic magnetic field and IBL
\cite{Kobayashi:2003kp}.

The other, perhaps more straightforward possibility is to consider an
additional component of GCR positrons with a {\it harder} source
spectrum that results in a harder propagated spectrum and therefore
leads to an increase in the positron fraction.

\subsection{Secondary accelerated electrons and positrons}
\label{sec:SecondarySource}

It has been suggested that {\it acceleration} of secondary $e^\pm$
produced through pp interactions inside the same sources where GCR
protons are accelerated, {\it e.g.}~SNRs, can produce a hard positron
component \cite{Blasi:2009hv}. We recapitulate here the essential
formalism of diffusive shock acceleration
\cite{Blandford:1987pw,Malkov:2001} which yields the spectrum of the
accelerated protons. This serves as the source term for calculating
the spectrum of the secondary $e^\pm$.

The phase space density, $f_\pm$, of secondary $e^-$ and $e^+$
produced by the primary GCR, both undergoing DSA, is described by the
steady state transport equation:
\begin{equation}
  u \frac{\partial f_{\pm}}{\partial x} =
  \frac{\partial}{\partial x} \left( D \frac{\partial}{\partial x} 
    f_\pm \right) + \frac{1}{3} \frac{\text{d} u}{\text{d} x} p 
  \frac{\partial f_{\pm}}{\partial p} + q_{\pm} \,,
\label{eqn:TransportEq}
\end{equation}
where $q_\pm$ is the source term determined by solving an analogous
equation for the primary GCR protons. (Ideally we should solve the
time-dependent equation, however we do not know the time-dependence of
the parameters and can extract only their effective values from
observations. This ought to be a good approximation for calculating {\em ratios} of secondaries to primaries from a large number of sources which are in different stages of evolution.) We consider
the usual setup in the rest-frame of the shock front (at $x=0$) where
$u_1$ ($u_2$) and $n_1$ ($n_2$) denote the upstream (downstream)
plasma velocity and density, respectively. The compression ratio of
the shock $r=u_1 / u_2 = n_2 /n_1$ determines the spectral index,
$\gamma = 3r/(r - 1)$, of the GCR primaries in momentum space (note
$\gamma = 2 + \Gamma$). To recover $\gamma \simeq 4.4$ as determined
from $\gamma$-ray observations (see Table \ref{tbl:GammaRays}) we set
$r \simeq 3.1$. As noted earlier the theoretical expectation is
however $r = 4$.

\begin{figure}[tb]
\centering
\includegraphics[width=0.45 \columnwidth]{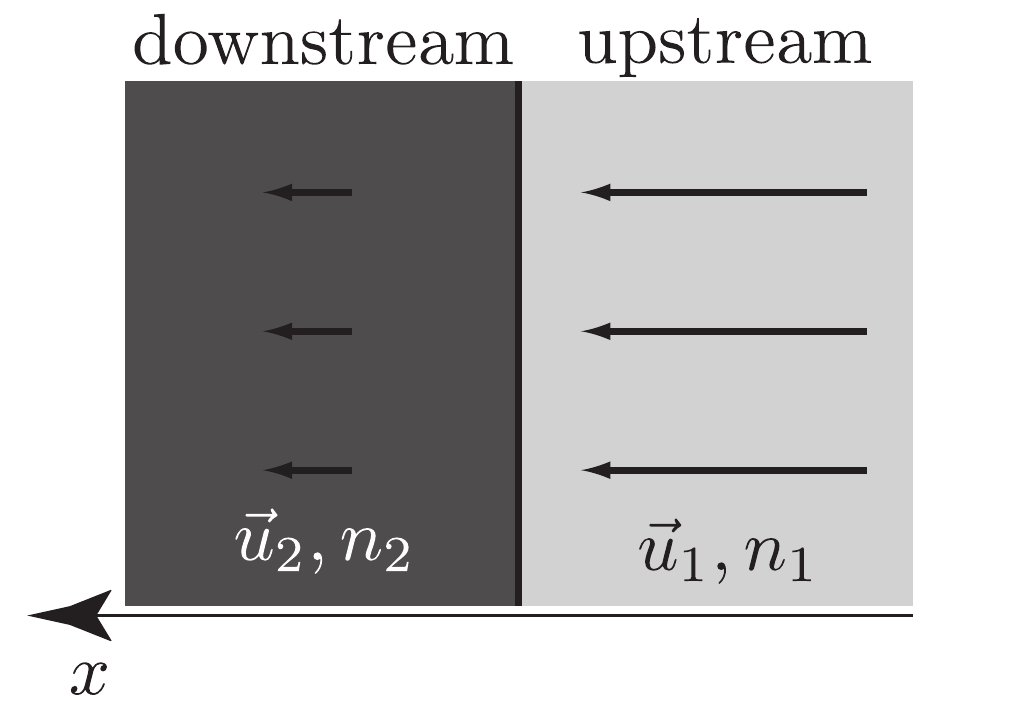}
\includegraphics[width=0.45 \columnwidth]{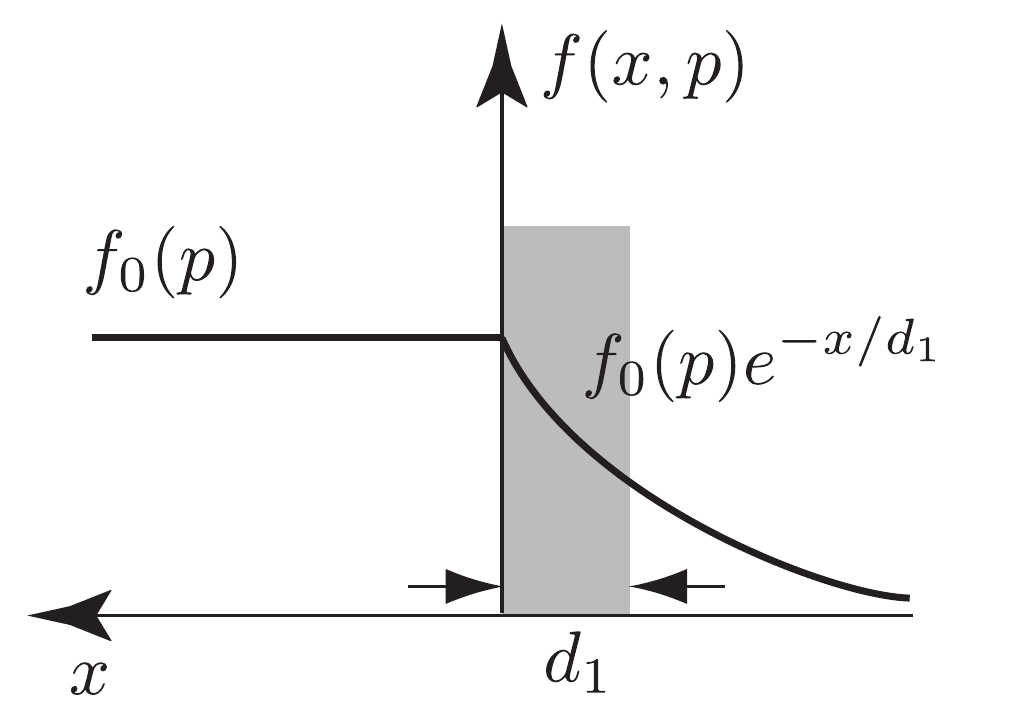}
\caption[]{DSA setup in the rest frame of the shock front. $u_1$
  ($u_2$) and $n_1$ ($n_2$) denote upstream (downstream) plasma
  velocity and density, respectively. The right panel shows the
  solution of the transport equation for the primary GCRs. Particles
  within a distance $D/u$ of the shock front participate in the
  acceleration process.}
\label{fig:DSA-d-02}
\end{figure}

For $x \neq 0$, Eq.~(\ref{eqn:TransportEq}) reduces to an ordinary
differential equation in $x$ that is easily solved taking into
account the spatial dependence of the source term
\begin{equation}
q_{\pm}^0(x,p) = \left \{
\begin{array}{ll}
q_{\pm, 1}^0 (p) \text{e}^{x \, u_1 / D (p_\text{p})} & \text{for } x < 0 ,\\
q_{\pm, 2}^0 (p) & \text{for } x > 0 ,\\
\end{array}
\right.
\label{eqn:injection}
\end{equation}
where the proton momentum $p_\text{p}$ should be distinguished from
the (smaller) momentum $p$ of the produced secondaries, the two being
related through the inelasticity of $e^\pm$ production: $\xi \simeq
1/20$. Assuming $D \propto p$ (Bohm diffusion) in the SNR, the
solution to the transport equation~(\ref{eqn:TransportEq}) across the
shock can then be written (see Appendix \ref{DSA}):
\begin{equation}
f_{\pm} = \left \{
\begin{array}{ll}
f_\pm^0 \text{e}^{x/d_1} -\frac{q_{\pm, 1}^0}{u_1}d_1 
\left(\frac{\text{e}^{x/d_1} - \text{e}^{\xi x/d_1}}{\xi - \xi^2}\right) 
& \text{for } x < 0, \\
f_\pm^0 + \frac{q_{\pm, 2}^0}{u_2} x & \text{for } x > 0, \\
\end{array}
\right.
\label{eqn:solution}
\end{equation}
where $d_1 \equiv D/u_1$ is the effective size of the region where
$e^-$ and $e^+$ participate in DSA (see Fig.~\ref{fig:DSA-d-02}).

The coefficients $f_\pm^0$ appearing in Eq.~(\ref{eqn:solution})
satisfy an ordinary differential equation dictated by
continuity across the shock front (see Appendix~\ref{DSA}). This has
the solution:
\begin{align}
  f_\pm^0 (p) = \gamma \left(\frac{1}{\xi} + r^2 \right)
  \int_0^p \frac{\text{d}p'}{p'} \left(\frac{p'}{p}\right)^\gamma
  \frac{D(p') q_{\pm, 1}(p')}{u_1^2}.
\label{eqn:fpm0}
\end{align}
Assuming Feynman scaling for the pp interaction, {\it i.e.} $p\,\text{d}
\sigma_\text{pp}/\text{d}p \propto \Sigma_\pm $ we can express the
momentum dependence of the source term as
\begin{align}\nonumber\label{eqn:defqpm}
  q_{\pm, 1}(p) &= \frac{c\,n_{\text{gas}, 1}}{4\pi p^2} \int_p^\infty
  \text{d}p'
  N_\text{CR} (p') \frac{\text{d}\sigma_{\text{pp} \to \text{e}^\pm + X}}{\text{d}p} \\
  &\simeq \frac{c\,n_{\text{gas},1}}{4\pi p^2} N_\text{CR} (p)
  \frac{\Sigma_\pm}{\gamma - 2} \,.
\end{align}
The maximum energy of protons is determined from the average maximum
$\gamma$-ray energy $E_{\text{max}} \simeq 20 \, \text{TeV}$
(see Table \ref{tbl:GammaRays}) through the inelasticity of the
$\text{pp} \rightarrow \gamma + X$ process as $\sim 20\,\text{TeV}/0.15
 \approx 100 \, \text{TeV}$ \cite{Aharonian:2006ws}.

We can easily interpret the solution (\ref{eqn:solution}) in terms of
power laws in momentum. The second term downstream, $(q_2^0/u_2) x$,
follows the spectrum of the primary GCRs ($\propto p^{-\gamma}$) and
describes the production of secondary $e^-$ and $e^+$ that are then
advected away from the shock front. However, secondaries that are
produced within a distance $\sim D/u$ from the shock front are subject
to DSA (see Eq.~\ref{eqn:injection} and Fig.~\ref{fig:DSA-d-02}). The
fraction of secondaries that enters the acceleration process is thus
given by the ratio of the relevant volumes, {\it i.e.}~$(D/u_1)/(u_2
\, \tau_{\text{SNR}})$, and the number density injected into the
acceleration process is $ (1/\xi + r^2 ) D q_{\pm, 1}/u_1^2$. This
rises with energy because of the momentum dependence of the diffusion
coefficient ($D(p) \propto p$) so the first term downstream in
Eq.~(\ref{eqn:solution}) gets harder: $f_\pm^0 (p) \propto p^{-\gamma
  + 1}$.

The injection spectrum $R_\pm$ is obtained by integrating the steady
state solution over the volume of the SNR:
\begin{equation}
R_\pm = 4\pi p^2 4\pi \int_0^{u_2 \tau_\text{SNR}} \text{d}x \, x^2 
f_\pm (x, p).
\end{equation}
The resulting source spectrum, $R_{\pm}$, is thus the sum of two power
laws,
\begin{equation}
  R_\pm \simeq R_\pm^0 \, p^{-\gamma + 2} \left[1 
  + \left(\frac{p}{p_\text{cross}}\right)\right] ,
\end{equation}
where the ``cross-over'' momentum, $p_\text{cross}$, satisfies
\begin{equation}
\label{eqn:pxdef}
D (p_\text{cross}) = \frac{3}{4}
 \frac{ru_1^2\tau_\text{SNR}}{\gamma(1/\xi + r^2)}\,.
\end{equation}
As has been noted \cite{Blasi:2009hv}, this mechanism is most
efficient for {\it old} SNRs where field amplification by the shock
wave is not very effective anymore. We therfore introduce a fudge
factor $K_\text{B}$ that parameterises the effect of the smaller field
amplification on the otherwise Bohm-like diffusion coefficient in the
SNR,
\begin{equation}
D (E) = 3.3 \times 10^{22} K_\text{B} \, \bigg(\frac{B}{\mu{\rm G}}\bigg)^{-1} 
\bigg(\frac{E}{\text{GeV}}\bigg) \, \text{cm}^2 \text{s}^{-1}.
\label{eqn:D(p)}
\end{equation}

\begin{figure}[t!]\centering
  \includegraphics[width=0.95\columnwidth]{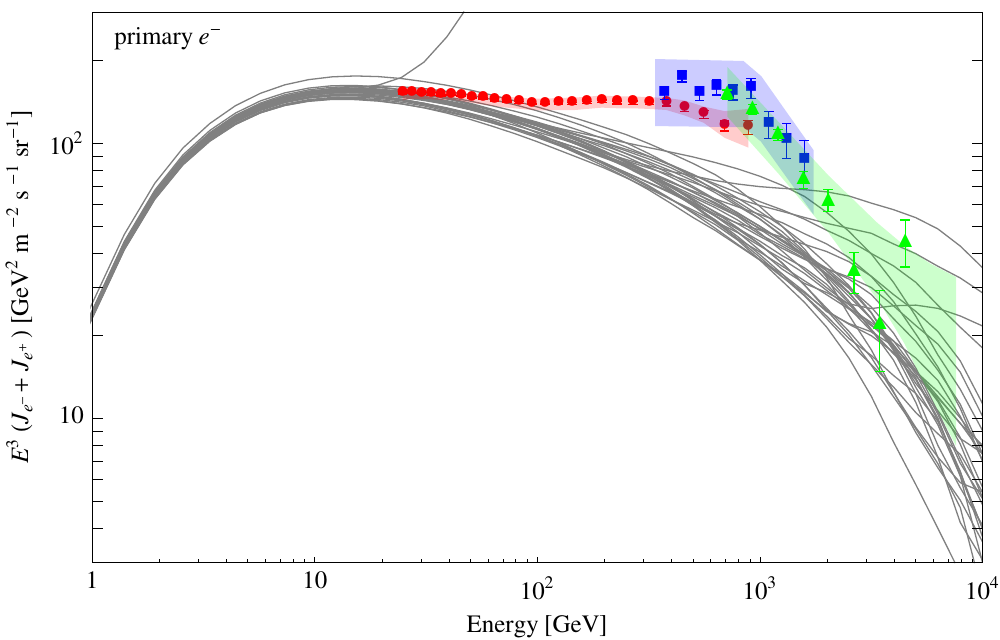}\\[0.3cm]
  \includegraphics[width=0.95\columnwidth]{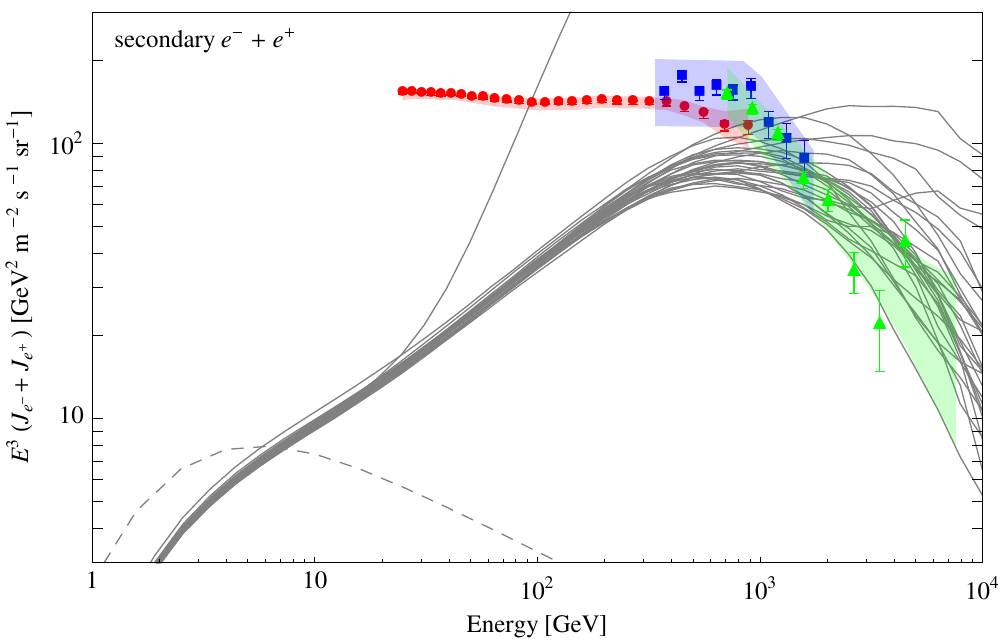}\\[0.3cm]
  \includegraphics[width=0.95\columnwidth]{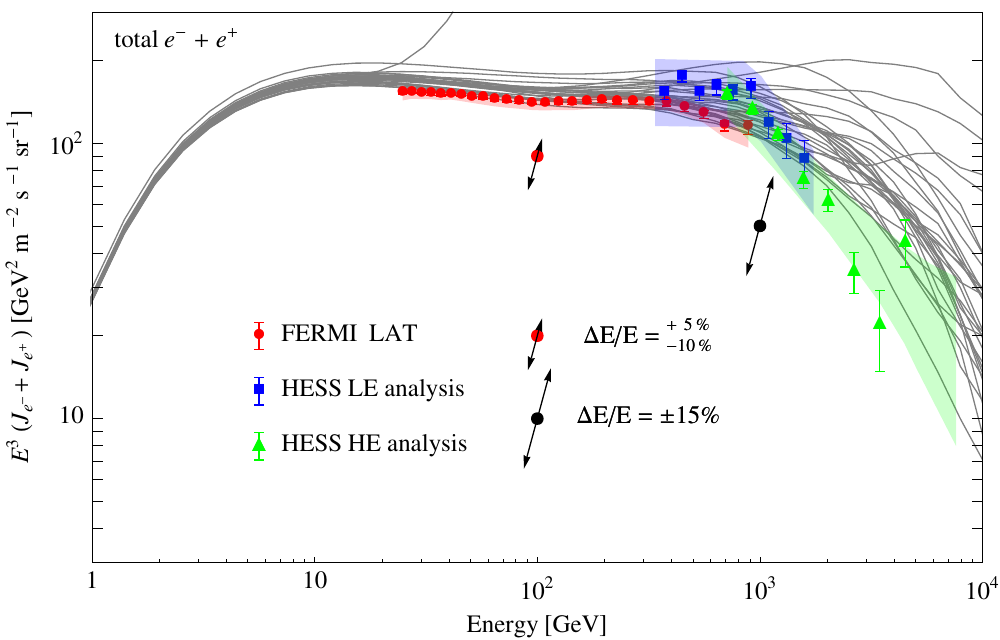}\\[0.3cm]
\caption{Predicted spectra of electrons and positrons with data from
  Fermi LAT \cite{Abdo:2009zk} (red circles) and HESS
  \cite{Collaboration:2008aaa, Aharonian:2009ah} (blue squares \&
  green triangles). The diagonal arrows show the energy scale uncertainty.
  {\bf Top:} Primary
  electrons after propagation to Earth. {\bf Middle:} Secondary
  electrons and positrons from cosmic ray interactions, created during
  propagation (dashed line) and created during acceleration in SNRs
  (full lines).  {\bf Bottom:} The sum of primary and secondary
  electrons and positrons.}
  \label{fig:E3J}
\end{figure}

The number of particles entering the acceleration process can of
course not exceed the total number of secondaries produced inside the
SNR. This effectively caps the growth of the term $D(p') q_{\pm, 1}^0
(p')/u_1^2$ once $(D/u_1)/(u_2 \, \tau_{\text{SNR}})$ becomes larger
than unity, a relation that defines a characteristic momentum scale
$p_\text{break}$. We therefore substitute in Eq.~(\ref{eqn:fpm0}),
\begin{equation}
\frac{D(p) q_{\pm, 1}^0 (p)}{u_1^2} \rightarrow \left \{
\begin{array}{ll}
\frac{D(p) q_{\pm, 1}^0 (p)}{u_1^2} & \quad \text{for } p < p_\text{break}, \\
\frac{D(p_\text{break}) q_{\pm, 1}^0 (p)}{u_1^2} & \quad \text{for } 
 p > p_\text{break}.
\end{array}
\right.
\end{equation}
The source spectrum $R_\pm$ thus returns to a $p^{-\gamma}$ dependence
around $p = p_\text{break}$. At even higher energies the secondary
spectrum cuts off at the same $E_\text{cut}$ as for primary electrons
(see sec. \ref{sec:PrimaryElectrons}).

Following Refs.~\cite{Blasi:2009hv, Blasi:2009bd}, the parameters are
chosen to be: $u_1 = 0.5 \times 10^8 \,\text{cm} \,\text{s}^{-1}$,
$n_{\text{gas}, 1} = 2 \,\text{cm}^{-3}$, $B = 1
\,\mu\text{G}$. Choosing $r = 3.1$ to recover $\Gamma = 2.4$ the
characteristic momenta $p_\text{cross}$ and $p_\text{break}$ turn out
to be,
\begin{align}\label{eqn:px}
p_\text{cross} &= 427 \, K_\text{B}^{-1} 
\left(\frac{\tau_{\text{SNR}}}{10^4 \, \text{yr}} \right) \, \text{GeV},\\
\label{eqn:pc}
p_\text{break} &= 7.7 \, K_\text{B}^{-1} \left( \frac{\tau_{\text{SNR}}}{10^4 \, 
\text{yr}} \right) \, \text{TeV}.
\end{align}

What is still missing is the normalization of the injection spectrum,
$R_{+}^0$, in the sources which is proportional to the normalisation
of the GCR protons, $N_{\text{GCR}}$, through
Eq.~(\ref{eqn:injection}). Usually a factor $K_{\text{ep}} \simeq
10^{-4} - 10^{-2}$ is introduced to normalize the electron component
with respect to the protons; this depends on how particles are
injected from the thermal background into the acceleration process and
is not reliably calculable from first principles. We can get around
this by assuming that the $\gamma$-rays detected from known SNRs by
HESS are of hadronic origin, as is expected in this framework. Thus we
can use the total luminosity of individual sources in $\gamma$-rays,
\begin{equation}
Q_{\gamma} =  4 \pi d^2 J_{\gamma} \, ,
\label{eqn:GammaLumi}
\end{equation}
to determine the normalization of the proton component and therefore
also the secondary injection rate $q_\pm^0$.

The compilation of $\gamma$ ray data on SNRs from HESS, see
Table~\ref{tbl:GammaRays}, suggests an average value $Q_\gamma^0
\simeq 5.7 \times 10^{33} \,\text{s}^{-1} \,\text{TeV}^{-1}$.  We find
then for the total spectrum
\begin{align}
R_+^0 = \tau_\text{SNR} Q_+^0 \simeq \tau_\text{SNR} 
 \frac{\Sigma_{+}}{\Sigma_\gamma} Q_\gamma^0 \,,
\end{align}
where $\Sigma_{+}$ (and analogously $\Sigma_\gamma$) are defined by
Eq.~(\ref{eqn:defqpm}), or explicitly
\begin{align}
R_+^0 = 7.4 \times 10^{48} \bigg(\frac{\tau_\text{SNR}}{10^4 \text{yr}}\bigg) 
\bigg(\frac{Q_\gamma^0}{5.7 \times 10^{33} \text{s}^{-1}\text{TeV}^{-1}}\bigg)  
\text{GeV}^{-1} .
\end{align}
In the Monte Carlo code we have explicitly input the experimentally
measured pp cross-section which gives a similar
normalisation as the estimate presented above assuming Feynman
scaling. The normalisation for secondary electrons is computed similarly.

The middle panel of Fig.~\ref{fig:E3J} shows an example of the flux of
secondary source $e^-$ and $e^+$ for $30$ ``histories'' of SNRs in our
Galaxy. Clearly this component can potentially match the high energy
Fermi LAT and HESS data.

We note that in our model, the contribution from secondary electrons
and positrons to the total flux is about twice as large as in
ref.\cite{Blasi:2009hv} where the primary injection spectrum was
assumed to be $\propto E^{-2}$, motivated by DSA theory. However
this is not consistent with $\gamma$-ray observations of SNRs as seen
from Table~\ref{tbl:GammaRays}.

\section{Results}
\label{sec:Results}

\begin{table}[t]
\begin{tabular}{ c | c | c}
\hline\hline 
\multicolumn{3}{c}{Diffusion Model}\\
\hline

$D_0$ 			& $10^{28}\,\text{cm}^2\,\text{s}^{-1}$  & \multirow{3}{*}{$\Bigg\}$ \begin{minipage}[c]{0.5\columnwidth} \flushleft from GCR nuclear\\ secondary-to-primary ratios\end{minipage}} \\
$\delta$			& $0.6$ &  \\
$L$				& $3$ 			 $\text{kpc}$ \\
$b$ 				& $10^{-16}\,\text{GeV}^{-1} \, \text{s}^{-1}$ &  CMB, IBL and $\vec{B}$ energy densities \\
\hline\hline
\multicolumn{3}{c}{Source Distribution}\\
\hline
$t_\text{max}$		& $1 \times 10^8\,\text{yr}$ 	& from $E_{\text{min}} \simeq 3.3 \, \text{GeV}$  \\
$\tau_{\text{SNR}}$ 	& $10^4\,\text{yr}$ 			& from observations \\
$N$				&$3 \times 10^6$		& from number of observed SNRs \bigstrut[b] \\
\hline\hline
\multicolumn{3}{c}{Source Model}\\
\hline
$R_{e^-}^0$ 		& $1.8 \times 10^{50}\,\text{GeV}^{-1}$ &fit to $e^{-}$ flux at $10 \, \text{GeV}$ \\
$\Gamma$		& $2.4$		&average $\gamma$-ray spectral index\\
$E_\text{max}$		&  $20\,\text{TeV}$				& typical $\gamma$-ray maximum energy \\
$E_{\text{cut}}$		&  $20\,\text{TeV}$				& DSA theory \\
$R_{+}^0$		& $7.4 \times 10^{48}\,\text{GeV}^{-1}$ & {\it cf.}~Sec.~\ref{sec:SecondarySource} 	\\
$K_\text{B}$ 		&  $15$  & only free parameter (for fixed $\Gamma$)\\
\hline \hline
\end{tabular}
\caption{Summary of parameters used in the Monte Carlo simulation, for an injection spectral index $\Gamma \simeq 2.4$.}
\label{tbl:Parameters}
\end{table}

The parameters used in the Monte Carlo are given in Table
\ref{tbl:Parameters}. For an assumed injection spectral index
$\Gamma$, the only free parameter is $p_\text{cross}$ ({\it
  cf.}~Eq.~(\ref{eqn:pxdef})) or, equivalently, the factor
$K_\text{B}$ ({\it cf.}~Eq.~(\ref{eqn:D(p)})) which is determined by
fitting the total flux of electrons and positrons to the Fermi LAT and
HESS data (see Fig.~\ref{fig:E3J}). Adopting $\Gamma = 2.4$, we find
good agreement for $K_\text{B} \simeq 15$, which corresponds to a
cross-over of the primary and (accelerated) secondary components at
$p_\text{cross} \simeq 28$~GeV and a spectral break at $p_\text{break}
\simeq 510$~GeV ({\it cf.}~Eqs.~(\ref{eqn:px}) and (\ref{eqn:pc})).

\begin{figure*}
\includegraphics[width=\textwidth]{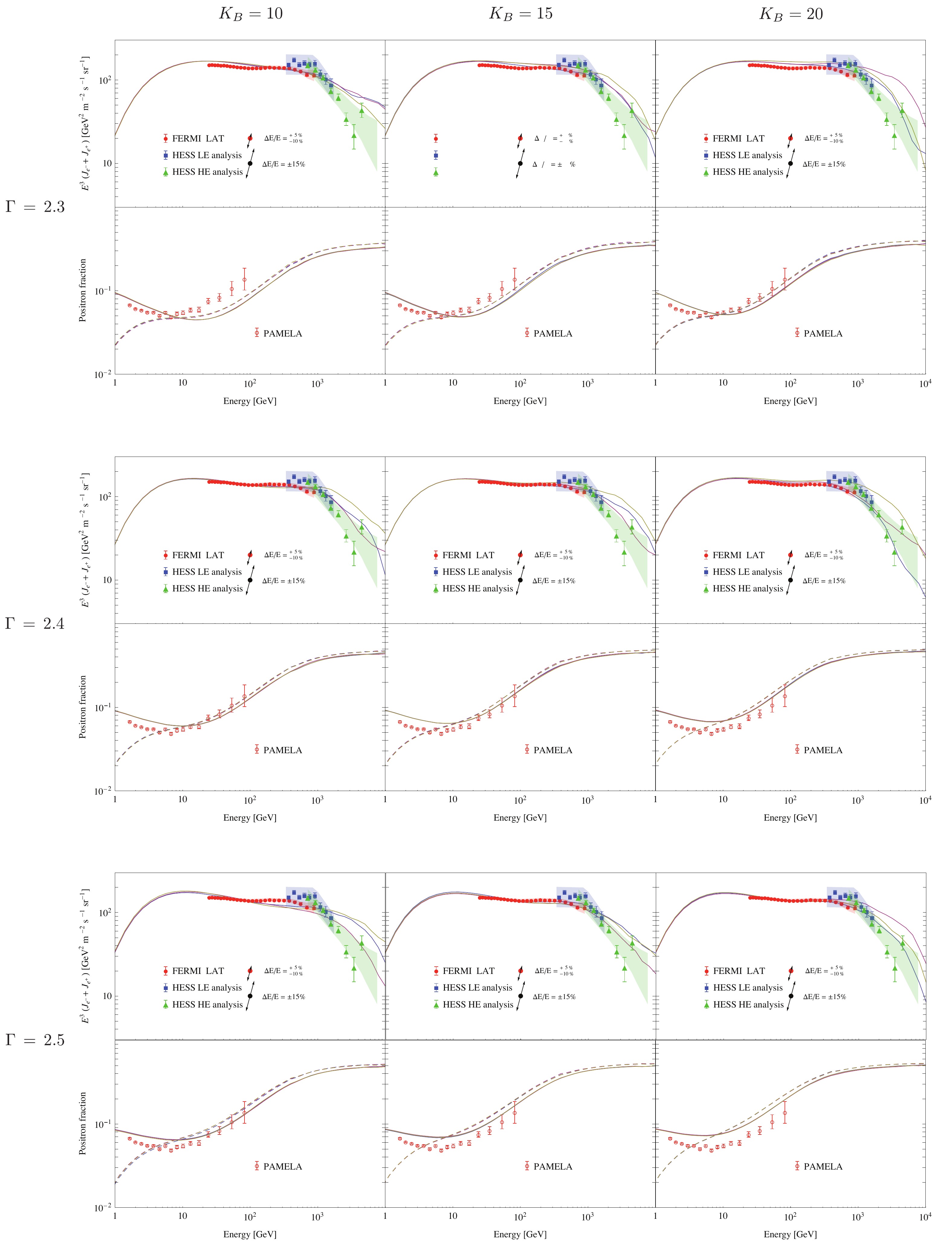}
\caption{The three best fits (out of 30 source ``histories'') to the
  total spectrum of electrons and positrons measured by Fermi
  LAT~\cite{Abdo:2009zk} (red circles) and
  HESS~\cite{Collaboration:2008aaa, Aharonian:2009ah} (blue squares \&
  green triangles), and the corresponding prediction for the positron
  fraction for different values of $\Gamma$ and $K_\text{B}$, for both
  charge-sign independent (full line) and charge-sign dependent
  (dashed line) Solar modulation (see text for details). The PAMELA
  data \cite{Adriani:2008zr} is shown for comparison (open red
  circles).}
  \label{fig:bestfit}
\end{figure*}

We have calculated the $\chi^2$ with respect to the combined Fermi LAT
and HESS data for each configuration $m$ of source distances and
times, $\{d_i, t_i \}_m$, over all energy bins $j$. The three best
``fits'' are shown in Fig.~\ref{fig:bestfit} for different values of
$K_\text{B}$ and for $\Gamma =$ 2.3, 2.4 and 2.5 (see
Table~\ref{tbl:Parameters}). The corresponding predictions for the
$e^+$ fraction are shown in the bottom panels.  These agree reasonably
well with the data down to 6~GeV; we would not expect agreement at
lower energies since we have neglected convection and reacceleration
during interstellar propagation. In fact the PAMELA measurements of
the $e^+$ fraction are systematically lower than previous
measurements, {\em e.g.} AMS-01 or HEAT, and it has been noted that
this discrepancy can be resolved by considering charge-sign {\em
  dependent} Solar modulation with $\phi_+ = 438 \, \text{MV}$ for
$e^+$ and $\phi_- = 2 \, \text{MV}$ for $e^-$ \cite{Gast:2009} (rather
than $\phi_+ = \phi_- = 600$~MV). This however seems to be at odds
with preliminary PAMELA data on the absolute electron flux
\cite{Mocchiutti:2009} which \textit{does} show substantial Solar
modulation. Accordingly in Fig.\ref{fig:bestfit} we have shown the
predicted $e^+$ fraction for both cases; note that this does not
affect our predictions for energies above 10 GeV.

Thus our fits to both the PAMELA and the Fermi LAT spectra, including
secondary $e^+$ accelerated in SNRs, provides a consistent picture of
current data on cosmic ray $e^-$ and $e^+$ between a few GeV and tens
of TeV. Turning the argument around, since a large fraction of the
$e^-$ and $e^+$ observed in GCR above hundreds of GeV are required to
be secondaries in this model, there \emph{must} be a large number of
hadronic cosmic ray accelerators in our Galaxy, some of which should
be quite nearby.

\begin{figure}[t]
\centering
\includegraphics[width=0.96\columnwidth]{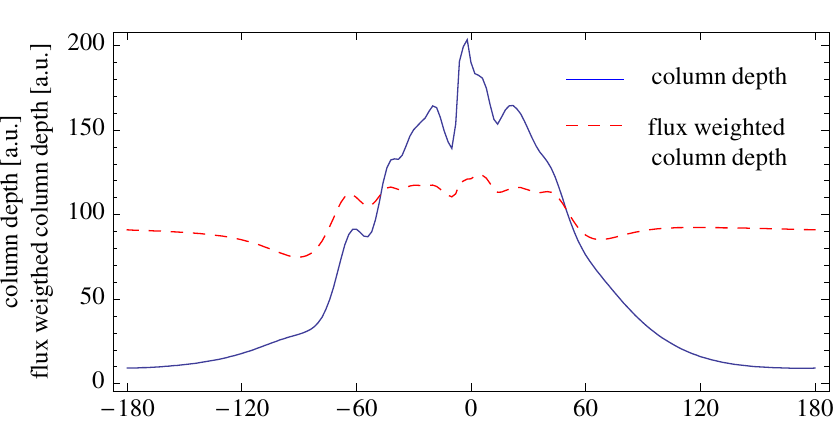}\\[0.4cm]
\hspace{0.3cm}\includegraphics[width=0.95\columnwidth]{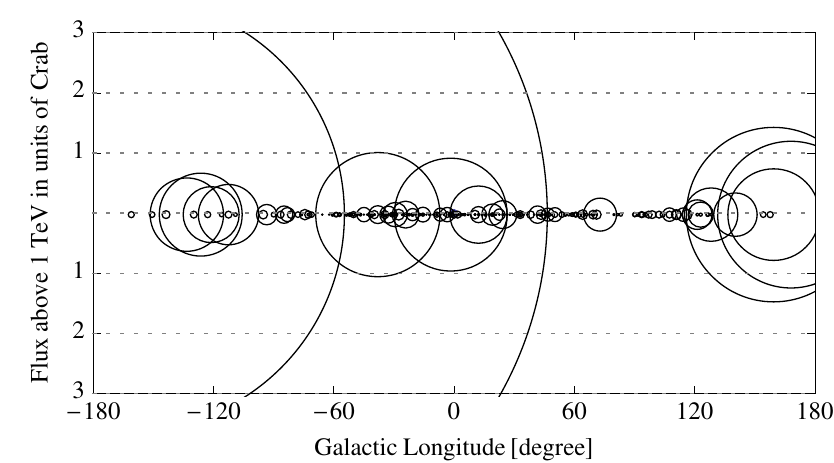}
\caption{{\bf Top:} The column depth and flux weighted column depth of
  the SNR density in the Galactic plane.  {\bf Bottom:} Example of a
  distribution of SNRs in $\gamma$-rays/neutrinos from the Monte Carlo
  simulation. The position of a circle denotes the Galactic longitude
  of the source and the radius is proportional to the brightness in
  units of the Crab nebula. One source whose circle exceeds the
  vertical scale is $\sim 500$~pc from Earth and has a total
  integrated flux above 1~TeV of $\sim 6$ times the Crab Nebula.}
  \label{fig:SNRcolumndepth}
\end{figure}

An independent test of the model is provided by the usual `messengers'
of such hadronic acceleration environments, namely $\gamma$-rays
and neutrinos. Taking the known distribution of SNRs in the Galaxy
(see Sec.~\ref{sec:SourceDistribution}) we have calculated the column
depth in SNRs in the Galactic disk as seen from Earth,
\begin{equation}
X (\phi') = \int_0^\infty \text{d}r' \, r' g (r (r', \phi'), \phi(r', \phi')),
\label{eqn:ColumnDepth}
\end{equation}
and show this in the top panel of Fig.~\ref{fig:SNRcolumndepth}. As
expected, the column depth is largest towards the Galactic
centre. However, the quantity that is more important for observations
is the brightness of sources. We have therefore weighted the integrand
in Eq.~(\ref{eqn:ColumnDepth}) by $1/r^2$ and this flux weighted
column depth is also shown in the top panel of
Fig.~\ref{fig:SNRcolumndepth}. We note that although the maximum
brightness is still expected around the Galactic Centre, the
brightness in other directions is smaller by only $\sim 30\%$ because
the sources in the closest spiral arms are then dominant (if they are
actually there of course).

This is illustrated in the bottom panel of
Fig.~\ref{fig:SNRcolumndepth} by an example distribution of SNRs from
the Monte Carlo simulation, denoted by circles. The position of the
circle denotes the Galactic longitude and the radius is proportional
to the brightness in units of the Crab Nebula, {\it i.e.} an
integrated flux of $(1.98 \pm 0.08) \times 10^{-11} \,\text{cm}^{-2}
\,\text{s}^{-1}$ above $1 \,\text{TeV}$ \cite{Aharonian:2004wa}. For a
source of luminosity of $Q_{\gamma}^0 = 5.7 \times 10^{33} \,
\text{TeV}^{-1} \,\text{s}^{-1}$ (see Table~\ref{tbl:GammaRays}) at
distance $d$, the integrated flux above 1~TeV is,
\begin{align}
F_\gamma (> 1\text{TeV}) &= \frac{1}{4\pi d^2} 
\int_{1 \text{TeV}} \text{d}E\, Q_\gamma \nonumber \\
&\simeq 8.5 \times 10^{-12} \left(\frac{d}{2 \,\text{kpc}}\right)^{-2} 
\,\text{cm}^{-2} \,\text{s}^{-1} \,,
\end{align}
{\it i.e.}~about 40 \% of the Crab Nebula flux at $d = 2$ kpc. It is
seen that although most of the sources are clustered towards the
Galactic centre, there are several bright sources at large longitudes
as well. We find typically $\sim 3$ sources brighter than the Crab (or
$\sim 7$ brighter than 50 \% Crab).

The adopted distribution of SNRs (Sec.~\ref{sec:SourceDistribution})
and the average luminosity per source determined from a compilation of
known sources (Table \ref{tbl:GammaRays}) thus leads to the prediction
of several nearby SNRs with fluxes of the order of the Crab
Nebula. Note, however, that close sources could be rather extended and
thus have escaped detection by HESS in one of its surveys of the Milky
Way \cite{Aharonian:2005kn, Chaves:2009zza, Chaves:2009bq}. For
example, a diameter of $\sim50$~pc which is a typical value for a very
old SNR, corresponds to $1.5^\circ$ at 2~kpc.

Extended $\gamma$-ray luminous SNRs can however be detected by MILAGRO
\cite{Abdo:2007ad} with its larger field of view. A survey in Galactic
longitude $l \in [30^\circ, 220^\circ]$ and latitude $b \in
[-10^\circ, 10^\circ]$ has revealed 6 new sources at a median energy
of 20~TeV, several of which are spatially extended. The flux from a
SNR of the above luminosity at $d = 2 \,\text{kpc}$ is
$Q_\gamma^0/(4\pi d^2) \simeq 1.2 \times 10^{-11} \,\text{TeV}^{-1}
\,\text{cm}^{-2} \,\text{s}^{-1}$ at 1 TeV. Scaled with a spectral
index of 2.4 to 20~TeV, this gives $Q_\gamma^0/(4\pi d^2) \, 20^{-2.4}
\simeq 9.0 \times 10^{-15} \,\text{TeV}^{-1} \,\text{cm}^{-2}
\,\text{s}^{-1}$ which is in the range of the unidentified MILAGRO
sources \cite{Abdo:2007ad}. We note that the MILAGRO source MGRO
J1908$+$06 was recently confirmed by HESS
\cite{deOnaWilhelmi:2009zza}, though with a smaller angular extent of
$\sim 0.7^\circ$. However, correlating unidentified MILAGRO sources
with the FERMI BSL list \cite{Abdo:2009ku, Abdo:2009mg} seems to
favour associations with pulsars, although several new unidentified
extended sources have also been found.

Hadronic sources of cosmic rays should also be visible by their
neutrino emission. On general grounds, the neutrino luminosity (from
$\pi^\pm$ decay) can be directly related to the $\gamma$-ray
luminosity (from $\pi^0$ decay) and should be of the same order of
magnitude since pp interactions produce $\pi^+$, $\pi^0$ and $\pi^-$
in roughly equal numbers. Each of the three neutrinos produced in the
decay chains $\pi^+\to\mu^+\nu_\mu\to e^+\nu_e\bar\nu_\mu\nu_\mu$ and
$\pi^-\to\mu^-\bar\nu_\mu\to e^-\bar\nu_e\nu_\mu\bar\nu_\mu$ carries
about half of the energy of each photon produced in the decay
$\pi^0\to\gamma\gamma$. Hence, the ratio of neutrinos to photons
produced on average is $\sim 3:1$ and the total neutrino luminosity is
\begin{align}\nonumber
  Q_{\text{all }\nu} (E_\nu) &\simeq 6Q_\gamma (2E_\nu)\\
  &\simeq 6 \times 2^{-\Gamma} Q_\gamma^0 
 \left(\frac{E_\nu}{\text{TeV}}\right)^{-\Gamma}.
\end{align}
Presently the largest cosmic neutrino detector is the IceCube
observatory~\cite{Ahrens:2003ix} under construction at the South
Pole. IceCube observes high energy neutrinos via their interactions
with nucleons in the vicinity of the detector and subsequent
\v{C}erenkov light emission of energetic charged particles in the
transparent glacial ice. The most important signal for neutrino
astronomy is the \v{C}erenkov radiation by muons produced via charged
current interactions of muon neutrinos. Since the muon inherits the
large boost of the initial neutrino the point source resolution is
$\sim 1^\circ$. The large background signal of atmospheric muons is
efficiently reduced for upward-going muons, {\it i.e.}~neutrino
sources which are somewhat below the horizon. Hence, IceCube is mainly
sensitive to neutrino point sources in the northern sky, which
excludes SNRs in the direction of the Galactic centre.

Neutrino emission associated with galactic TeV $\gamma$-ray sources
has been investigated by many
authors~\cite{Anchordoqui:2006pb,Kistler:2006hp,Beacom:2007yu,Halzen:2007ah,Halzen:2008zj,GonzalezGarcia:2009jc}
including also the HESS sources used in our analysis. In particular,
Ref.~\cite{Kistler:2006hp} investigates the prospects of neutrino
detection for the SNRs HESS J0852.0--463, J1713--381, J1804--216,
J1834--087 (see Table~\ref{tbl:GammaRays}) in the proposed KM3NeT
detector in the Mediterranean which will see the Galactic centre
region. The muon neutrino rate is expected to be a few events per year
for such sources.

Due to flavour oscillations of neutrinos with large mixing angles, the
initial flavour composition $Q_{\nu_e}:Q_{\nu_\mu}:Q_{\nu_\tau} \simeq
1:2:0$ from pion decay is expected to become $\sim 1:1:1$ at
Earth. The TeV muon neutrino point flux from a hadronic $\gamma$-ray
source located at a distance $d$ and with a power-law index $\Gamma
\simeq 2.4$ is thus $F_{\nu_\mu}(>1~\text{TeV})\simeq 2^{1 - \Gamma}
F_{\gamma}(>1~\text{TeV})$, hence
\begin{align}
F_{\nu_\mu}(>1~\text{TeV}) \simeq 3.2 \times 10^{-12} 
\left(\frac{d}{2~\text{kpc}} \right)^{-2} \, \text{cm}^{-2} \, \text{s}^{-1}.
\end{align}
This should be compared to the results of searches for neutrino point
sources in the northern sky, in particular the close-by SNR Cassiopeia
A (see Table~\ref{tbl:GammaRays}), using data taken with AMANDA-II
(the predecessor of IceCube) during 2000--2006~\cite{Abbasi:2008ih}
and, more recently, with the first 22 strings of IceCube during
2007--08~\cite{Abbasi:2009iv}. The average 90\% C.L.~upper limit on
the integrated $\nu_\mu$ flux in the energy range 3~TeV to 3~PeV
is~\cite{Abbasi:2009iv}
\begin{equation}
F_{\nu_\mu} \leq 4.7\times10^{-12}~{\rm cm}^{-2}\,{\rm s}^{-1},
\end{equation}
{\it i.e.} well above the flux of $\sim7\times10^{-13}~{\rm cm}^{-2}\,{\rm s}^{-1}$ expected from a SNR at 2~kpc, assuming $\Gamma=2.4$.

The full 80 string configuration of IceCube thus has excellent 
prospects to identify these SNRs. A point source in the
northern sky with an $E^{-2}$ muon neutrino flux,
\begin{equation}
\label{eqn:icecubesen}
F_{\nu_\mu} \simeq 7.2 \times 10^{-12} \text{cm}^{-2}\,\text{s}^{-1},
\end{equation}
in the TeV-PeV range can be detected with a $5\sigma$ significance
after three years of observation. This does depend somewhat on the
spectral index and energy cut-off, since the signal (after ``level 2
cuts'') peaks at an energy of $\sim10$~TeV~\cite{Ahrens:2003ix}. As
mentioned previously, our analysis predicts on average $\sim 3$ nearby
$\gamma$-ray sources stronger than Crab with corresponding muon
neutrino fluxes larger than $\sim 7 \times
10^{-12}\,\text{cm}^{-2}\,\text{s}^{-1}$. Note that although the
Galactic centre is not in the field of view of IceCube, SNRs following
the spiral arm structure of the Galaxy are expected to be detected
also in the Galactic anti-centre direction, as seen in the example
distribution shown in the bottom panel of
Fig.~\ref{fig:SNRcolumndepth}.

\section{Summary}
\label{sec:DiscussionSummary}

Supernova remnants have long been suspected to be the sources of
Galactic cosmic rays. We have discussed a recent
proposal~\cite{Blasi:2009hv} that proton-proton interactions in the
shocks of SNRs followed by the diffusive shock acceleration of the
secondary positrons produced can flatten the spectrum of the
secondaries relative to that of the primaries. These hard spectra may
be the origin of the recently observed cosmic ray ``excesses'' ---
both the $e^+$ fraction observed by PAMELA \cite{Adriani:2008zr} and
the $e^{-} + e^{+}$ flux measured by Fermi LAT \cite{Abdo:2009zk} and
HESS \cite{Collaboration:2008aaa,Aharonian:2009ah}.

We have investigated how $\gamma$-ray emission of SNRs -- assumed to
be of the same hadronic origin as the positrons -- together with
cosmic ray data, constrain the acceleration of positrons. We have
accounted for the spatial and temporal discreteness of SNRs via a
Monte Carlo exercise, drawing samples from a realistic galactic
distribution with the observed SN rate. For the diffusion parameters
we have adopted standard values derived from cosmic ray
nuclear-to-primary ratios, as well as the energy densities of galactic
radiation and magnetic fields.

We have compiled a list of all $\gamma$-ray emitting SNRs observed by
HESS and determined the mean value of the flux, which fixes the
hadronic interaction rate in the SNR. Low energy data from PAMELA on
the absolute $e^-$ flux was used to normalize the primary flux of
$e^-$. The contribution from accelerated $e^+$ was then found by
fitting the $e^- + e^+$ flux to Fermi LAT and HESS data, adjusting the
(only) free parameter $K_\text{B}$ which determines the diffusion rate
near SNR shocks. 

\begin{figure}[b]
\centering
\includegraphics[width=0.96\columnwidth]{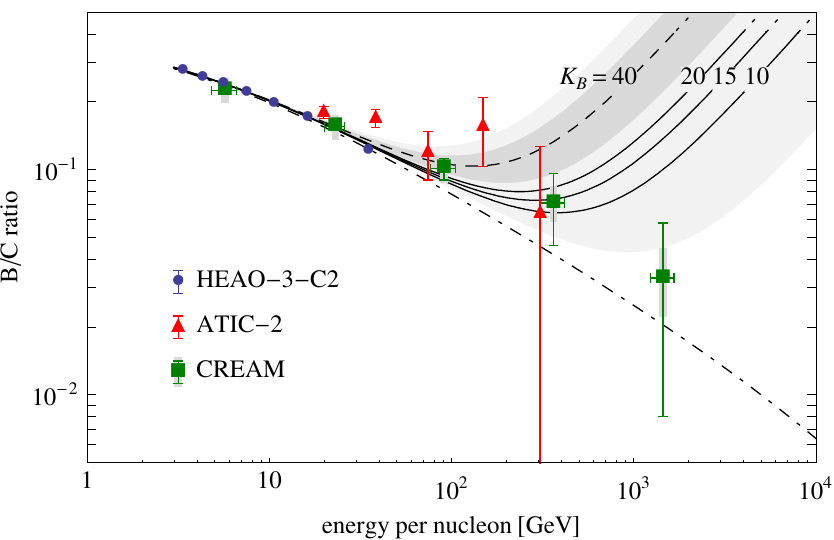}\\[0.4cm]
\caption{The B/C ratio in cosmic rays along with model predictions
  (after Ref. \cite{Mertsch:2009ph}) --- the leaky box model with
  production of secondaries during propagation only (dot-dashed line),
  and including production and acceleration of secondaries in a nearby
  SNR (solid lines) for values of the diffusion coefficient near the
  shock wave which best fit the $e^{\pm}$ spectrum (see
  Fig.\ref{fig:bestfit}). The dashed line corresponds to the value of
  the diffusion co-efficient required to fit the ATIC-2 data on Ti/Fe
  (from Ref.\cite{Mertsch:2009ph}), along with the $1 \sigma$ and $2
  \sigma$ error bands. The data points are from HEAO-3-C2 (circles)
  \cite{Engelmann:1990aa}, ATIC-2 (triangles) \cite{Panov:2007fe} and
  CREAM (squares) \cite{Ahn:2008my}.}
  \label{fig:B2C}
\end{figure}

The spectra of $e^+$ and $e^-$ thus derived agrees well with the $e^+$
fraction observed by PAMELA in the range $5-100$~GeV. The apparent
deficit at lower energies can be attributed to convection and
diffusive reacceleration of primary electrons that become important at
these energies and were neglected in our analysis. The flux of $e^+$
and $e^-$ becomes dominated by the accelerated secondary component at
high energies; the corresponding $e^+$ fraction levels out at
$\sim0.4$, reflecting the relative multiplicity of $e^+$ and $e^-$
produced by p-p interactions.

We note that although the value of $\sim 10-20$ found here for
$K_\text{B}$ differs from the value of $\sim 40$ we had found earlier
from our analysis of the titanium-to-iron (Ti/Fe) ratio in cosmic rays
\cite{Mertsch:2009ph}, the latter determination is subject to large
uncertainties due to poor experimental statistics at high
energies. This affects our previous prediction \cite{Mertsch:2009ph}
for the boron-to-carbon (B/C) ratio and we show in Fig. \ref{fig:B2C}
the corresponding $1 \sigma$ and $2 \sigma$ error bands, along with
the prediction taking $K_\text{B} \sim 10-20$ as indicated by our new
fits to the $e^\pm$ spectrum. The predicted upturn at
high energies will soon be tested by PAMELA and AMS-02.

To be consistent with our overall framework the $\gamma$-rays observed
from SNRs have been assumed to be of hadronic origin. The known
spatial distribution of SNRs then implies (on average) several nearby
sources with a $\gamma$-ray flux comparable to the Crab. We have
speculated that some unidentified MILAGRO sources \cite{Abdo:2007ad}
might correspond to such old SNRs. Moreover, the same hadronic
processes in SNRs will inevitably produce high energy neutrinos which
can be detected in cubic-km telescopes such as
IceCube~\cite{Ahrens:2003ix}. The neutrino luminosity can be directly
related to the $\gamma$-rays and is not connected to the hypothetical
acceleration of $e^+$ and $e^-$ in the sources as in our present
model. Nevertheless, similarly to the previous argument, we expect on
average a few nearby sources, some of which may may also lie within
the field of view of IceCube and can thus be detected with high
statistical significance after three years of data taking.

While our calculational framework is based on first-order Fermi
acceleration by SNR shock waves, we have noted that in detail the
observations do not fit the theoretical expectations, {\it e.g.,}~the
shock compression ratio inferred from the observed $\gamma$-ray
spectrum ($\sim E^{-2.4}$) is 3.1 rather than 4 as is expected for a
strong shock \cite{Blandford:1987pw}. Going beyond the test particle
approximation, the generic expectation in such a process is for
particle spectra which are much flatter than those observed ($\sim
E^{-1.4}$ and slightly {\it concave}), when the back reaction of the
cosmic rays on the shock is taken into account \cite{Malkov:2001}. By
contrast, the observed radio spectrum of Cassiopeia A is slightly
\emph{convex} and this, as well as the morphology and time evolution
of radio emission from such young SNRs, can be well explained in terms
of \emph{second-order} Fermi acceleration by plasma turbulence behind
the shock wave \cite{CowsikSarkar:1984}. Moreover the observed spatial
correlation between the $\gamma$-ray emission and the hard X-ray
emission from some SNRs argues for a leptonic rather than hadronic
origin and further observations are necessary to resolve this issue
\cite{Pohl:2008pq}. It has been argued that cosmic ray protons and
nuclei may well have different sources (e.g. ``superbubbles'' formed
by multiple supernovae) than the cosmic ray electrons
\cite{Butt:2009}.  The additional predictions made in this paper
concerning the visibility of hadronic accelerators in $\gamma$-rays
and neutrinos, tied to the expectations for the fluxes of the
accelerated {\it secondary} positrons in cosmic rays, will hopefully
enable further consistency tests of the SNR origin hypothesis for
galactic cosmic rays.

\section*{Acknowledgements}

We thank Tyce DeYoung, Pascuale Serpico and Andrew Strong for useful
discussions. This work was supported by the EU Marie Curie Network
``UniverseNet'' (HPRN-CT-2006-035863) and STFC (PP/D00036X/1).

\appendix

\section{Green's Function of Diffusion Equation}\label{Green}

We solve Eq.~(\ref{eqn:diffusion}) for a halo of extent $\pm L$ in $z$
direction, neglecting the boundaries in the radial direction. The
Green's function for the flux of electrons from a source at $\vec{r}$
that went off a time $t$ ago with a spectrum $Q(E)$, is:
\begin{align}
\label{Greensfunction}
& G_\text{disk}(E, \vec{r}, t) \nonumber \\
=& \sum_{n=-\infty}^\infty \frac{1}{(\pi\ell^2)^\frac{3}{2}}
\text{e}^{-\vec{r}_n^2/\ell^2}
Q \left(\frac{E}{1-b_0 E t}\right) \left(\frac{1}{1 - b_0 E t}\right)^2 \nonumber \\
=& \frac{1}{\pi\ell^2} \text{e}^{-\vec{r}_{\parallel}^2/\ell^2} 
\frac{QE}{(1 - b_0 E t)^3} \frac{1}{\ensuremath{\ell_\text{cr}}} \chi
\left(z/\ell_\text{cr}, \ell/\ell_\text{cr}\right),
\end{align}
where
\begin{align}
\label{functionchi}
\chi (\hat{z}, \hat{\ell}) &\equiv \frac{1}{\sqrt{\pi} \hat{\ell}} 
\sum_{n=-\infty}^\infty \text{e}^{-\hat{z}_n^2/\hat{\ell}^2},
\end{align}
and the diffusion length $\ell$ is defined as
\begin{align}
\label{difflength}\nonumber
\ell^2 &= 4 \int_E^{E/(1 - b(E)t)} \text{d}E' \, \frac{D(E')}{b(E')} \\
&= \frac{4 D_0}{b_0 (1 - \delta)} 
\left[E^{\delta - 1} - \left(\frac{E}{1-b_0 E t}  \right)^{\delta - 1}\right],
\end{align}
with $\ensuremath{\ell_\text{cr}} \equiv 4 L/\pi$.
If we neglect the spatial extent of the disk and 
set $z = 0$, the function $\chi(0,
\hat{\ell})$ is approximately:
\begin{equation}
\chi (0 , \hat{l}) \simeq \left\{\begin{array}{ll} 
\frac{4}{\pi} e^{-\hat{l}^2} & \text{for } \hat{l} \gg \frac{\pi}{4}, \\ 
\frac{1}{\sqrt{\pi} \hat{l}} & \text{for } \hat{l} \ll \frac{\pi}{4}.
\end{array} \right.
\end{equation}
In practice both limits can be connected at $\hat{l} \simeq 0.66$
such that the approximated $\chi(0, \hat{\ell})$ has a relative error
of at most 0.5\%. 
We motivate the choice of the parameters of our
diffusion model from an analysis of nuclear secondary-to-primary
ratios~\cite{Strong:2007nh}: $D_0 = 10^{28} \,\text{cm}^2
\text{s}^{-1}$, $\delta = 0.6$, $L = 3$~kpc, and from the Galactic
magnetic field and interstellar radiation fields
\cite{Kobayashi:2003kp}: $b_0 = 10^{-16} \, \text{GeV}^{-1} \,
\text{s}^{-1}$.

\section{Boundary Conditions of DSA}
\label{DSA}

The solution~(\ref{eqn:solution}) of 
Eq.~\ref{eqn:TransportEq} satisfies 
\begin{equation}
\lim\limits_{x \to -\infty}f_\pm = 0 \, ,  
\lim\limits_{x \to -\infty}\frac{\partial f_\pm}{\partial x} = 0\,\,
\text{and}\,\,\left| \lim\limits_{x \to \infty} f_\pm \right| < \infty \,,
\end{equation}
Continuity at the shock front $x=0$ requires:
%
\begin{equation}
\left. D \frac{\partial f_\pm}{\partial x} \right|_{x=0^-} -\left. 
D \frac{\partial f_\pm}{\partial x}\right|_{x=0^+} 
=
\frac{1}{3} (u_2 - u_1) p \frac{\partial f_\pm^0}{\partial p} \,,
\end{equation}
yielding the differential equation,
\begin{equation}
  p \frac{\partial f_\pm^0}{\partial p} = -\gamma f_\pm^0 
  + \gamma \left(\frac{1}{\xi} + r^2\right) \frac{D q_1^0}{u_1^2} \, .
\end{equation}
This is readily integrated with boundary condition $f^0_\pm (0) = 0$
and yields the $p$ dependence in Eq.~(\ref{eqn:fpm0}).


\end{document}